# A two-streams discrete ordinates solution of radiative transfer in 1-D spherical geometry

Charles H Aboughantous

*Louisiana State University, Department of Physics and Astronomy, Baton Rouge, Louisiana 70803, USA*

**Abstract.** The analytic solution for the two-streams discrete ordinates equations is carried out on the spatial domain in one-dimensional spherical geometry. The solution satisfies all the physical and the geometric conditions, namely: the natural boundary condition at the surface of the sphere, the symmetry condition at the center and the reciprocity of path. The solution obeys the inverse square law and is finite everywhere in the sphere except at the center. Graphical and numerical data obtained in cold absorbing spheres confirmed the analytical predictions. The conservation of energy is satisfied with high accuracy within round-off errors.

## 1. Prolog

We developed a revised version of the discrete ordinates method in one-dimensional spherical geometry in a previous paper [1]. The set of discrete ordinates, labeled $S_{2N}$, is structured from un-normalized circular functions and is intended to approximate the angular derivative of the transfer equation in conservation form. The resulting set of discrete ordinates equations is closed and conservative. We obtained the analytic solution of the centrifugal transfer equation for monochromatic radiation and used it to validate the proposed $S_{2N}$ in vacuum and in pure absorbing media. That solution is exclusive to centrifugal flow of energy. This paper is the natural extension to the previous one and is intended to validate $S_{2N}$ in a two-streams radiation field. Previous efforts to improve on the discretization of the angular derivative retained the same structure of the original angular parameters [2]. Recent efforts improved on the representation of the angular derivative and constrained the solution to agree with the analytic solution of the integral equation [3]. We hope that our efforts in this paper resolve the intrigue of the discretization of the angular derivative of the transfer equation in conservation form.

## 2. The transfer equations

Consider a sphere of radius $R$ hosting a concentric pellet of radius $\varepsilon$ and characterized by its opacity $\kappa$ and blackbody constant $B$. The sphere is assumed an isothermal gray medium, without scattering, immersed in a uniform and isotropic radiation field its intensity at the surface is $\psi_R^-$. The pellet could be arbitrarily small, or arbitrarily large subject to the condition $\varepsilon < R$, and it may be a cavity or, when justified, a gray isothermal medium. The solutions are developed for homogeneous media; it is shown in a later section that only minor changes are needed to apply the solutions to heterogeneous media. These limitations are sufficient to our purpose and do not impede the generality of the approach to obtain rigorous solutions with less cumbersome expressions. The radiation field in this setting is defined on $\mu \in [-1, +1]$ everywhere in the sphere and, when justified, in the pellet. The linearized transport equation for monochromatic radiation may be written in conservation form as:

$$\frac{\mu}{r^2}\partial_r\left(r^2\psi_r\right)+\frac{1}{r}\partial_\mu\left(\eta^2\psi_r\right)+\kappa\psi_r = \kappa B \qquad r \in [\varepsilon, R] \qquad (2.1)$$

where $\psi_r$ is the specific intensity in the direction specified by $\mu$, $\eta = (1-\mu^2)^{1/2}$ and $\partial_r$ is the tensor notation for the derivative with respect to $r$.

The equation (2.1) defines a boundary value problem subject to the natural boundary condition $\psi_R^-$ at the surface of the sphere, and symmetry boundary condition at the center. No other boundary condition at the center is permitted because (2.1) is not valid at the center. Instead, a specular reflection boundary condition is imposed at the surface of the pellet. *Albedo* boundary condition may also be imposed at the surface of the pellet but its validity is tied to the validity of the transfer equation in the close proximity of and at the center of the sphere.

The two-streams representation of (2.1) in $S_{2N}$ can be cast in the form [1]:

*Centrifugal equation* $\qquad \partial_r\psi_n^+ + \left[\frac{2+\beta_n^n}{r}+\lambda_n\right]\psi_n^+ - \frac{1}{r}\beta_n^{n-1}\psi_{n-1}^+ = \lambda_n B \qquad (2.2)$

*Centripetal equation* $\qquad \partial_r\psi_n^- + \left[\frac{2+\beta_n^n}{r}-\lambda_n\right]\psi_n^- - \frac{1}{r}\beta_n^{n-1}\psi_{n-1}^- = -\lambda_n B \qquad (2.3)$

where $\beta_n^n = \eta_n^2/w_n\mu_n$, $\beta_n^{n-1} = \eta_{n-1}^2/w_n\mu_n$ and $\lambda_n = \kappa/\mu_n$; the subscript $r$ in $\psi$ is dropped for notational convenience. It is replaced by the subscript $n$ for the $n$th direction on the set $\{\mu_n > 0\}$.

It is apparent that the two-streams equations differ in the sign of $\lambda$. It follows that in vacuum, the solutions of the two equations are identical: $\psi_{n,vac}^- = \psi_{n,vac}^+$, $\forall r$. This identity expresses the reciprocity of path that will be used to validate the proposed solution. The mathematical boundary condition for (2.2) and (2.3) are $\psi_\varepsilon^\pm$ at the surface of the pellet. These boundary values are not known *a priori*. They will be resolved in terms of the natural boundary condition $\psi_R^-$ by appropriate inversion of the solution of the centripetal equation.

The bracketed term of (2.2) is always positive. It was shown that in the case of one-stream radiation field the intensity function $\psi_n^+$ is not monotonic on $[\varepsilon, R]$ but exhibits a maximum close to the $\varepsilon$-surface [1]. This may not necessarily be true in the interior of the sphere considering the effect of the superposition of two opposing streams of radiation flow. By contrast, the bracketed term of (2.3) vanishes at some $r_n$ associated with a subset $\{\lambda_n\}$, suggesting that the spatial derivative could vanish at $r_n$. When this happens the spatial position $r_n$ could be either a point of inflection of the intensity function $\psi_n^-$, or an extremum. In the latter case, the function $\psi_n^-$ is not monotonic. Consequently, the total intensity would become non-monotonic as well whether or not there is blackbody emission.

*2.1. The formal mathematical solutions*

The solution for the centrifugal equation was carried out for the vector $\mathbf{\Psi}_r^+ = [\psi_1^+ \ \psi_2^+ \ ... \ \psi_N^+]^T$. It was cast in the form [1]:

$$\Psi_r^+ = \zeta^2 \mathbf{S}_{\varepsilon:r} \Psi_\varepsilon^+ + \mathbf{Q}_{r:\varepsilon}^+ \lambda \tag{2.4}$$

$$\mathbf{S}_{\varepsilon:r} = \mathbf{V}_{\varepsilon:r} \left[ \zeta^{\beta_n^n} e^{-\lambda_n(r-\varepsilon)} \right] \mathbf{V}_{\varepsilon:r}^{-1} \tag{2.5}$$

$$\mathbf{Q}_{r:\varepsilon}^+ = B \int_\varepsilon^r (x/r)^2 \mathbf{S}_{x:r} \, dx \tag{2.6}$$

The disposition of the double indices mirrors the disposition of the same in the ratio $\zeta = \varepsilon/r$ in the spherical operator $\mathbf{S}$, and the limits of integration of the emission term $\mathbf{Q}$. The eigentable $\mathbf{V}_{\varepsilon:r}$ is a lower triangular matrix its diagonal elements are $V_n^n = 1$, and the non-diagonal elements are obtained recursively from:

$$V_n^k = \frac{\Gamma_n^{n-1}}{\Gamma_k^k - \Gamma_n^n} V_{n-1}^k \qquad \text{for } k < n = 2, 3, ..., N \tag{2.7}$$

$$\Gamma_n^n = (2 + \beta_n^n) \ln \zeta - \lambda_n(r - \varepsilon)$$

$$\Gamma_n^{n-1} = -\beta_n^{n-1} \ln \zeta$$

Considering that (2.2) and (2.3) differ only in the sing of $\lambda_n$, the solution for the centripetal equation can be obtained directly from (2.4) by changing the sing of $\lambda_n$:

$$\Psi_r^- = \zeta^2 \mathbf{Z}_{\varepsilon:r} \Psi_\varepsilon^- - \mathbf{Q}_{r:\varepsilon}^- \lambda \tag{2.8}$$

$$\mathbf{Z}_{\varepsilon:r} = \mathbf{U}_{\varepsilon:r} \left[ \zeta^{\beta_n^n} e^{\lambda_n(r-\varepsilon)} \right] \mathbf{U}_{\varepsilon:r}^{-1} \tag{2.9}$$

$$\mathbf{Q}_{r:\varepsilon}^- = B \int_\varepsilon^r (x/r)^2 \mathbf{Z}_{x:r} \, dx \tag{2.10}$$

The eigentable $\mathbf{U}_{\varepsilon:r}$ is constructed with the recursion relation (2.7) simply by changing the sign of $\lambda$ in $\Gamma$. A useful property of the eigentables in vacuum is that they are independent of $r$ and $\varepsilon$, $\forall \{\varepsilon, r\}$:

$$\mathbf{U}_{\varepsilon:r}^{vac} = \mathbf{V}_{\varepsilon:r}^{vac} = \lim_{\varepsilon \to 0} \mathbf{U}_{\varepsilon:r} = \lim_{\varepsilon \to 0} \mathbf{V}_{\varepsilon:r} = \mathbf{V}$$

## 2.2. The centripetal intensity with natural boundary condition

In order to complete the solution for the centrifugal intensity (2.4) we seek the reflected centripetal vector at the surface of the pellet by first solving (2.8) for $\Psi_\varepsilon^-$ assuming $\Psi_r^-$ is known. We accomplish this by taking the *reciprocal* of the $\mathbf{Z}$-matrix of (2.9).

Historically, reciprocal of a matrix and inverse of a matrix meant the same thing [4]. Modern literature rarely uses the terminology reciprocal of a matrix, if at all. In this paper, we designate a *reciprocal* of a matrix to be the inverse of the matrix associated with symmetry application on the coupling parameters. This is understood if we realize that the coupling of the intensities at $r$ relative to $\varepsilon$ is symmetric to the coupling of the intensities at $\varepsilon$ relative to $r$. It is difficult to determine explicitly how such a coupling is achieved in the structure of $\mathbf{Z}$-matrix now that the derivative sprawls all over the matrix as $\beta$-parameters.

Define an operator $\mathbf{G}$ such that $(\mathbf{Z})_{reciprocal} \equiv \mathbf{G}\mathbf{Z}^{-1}$ and $\mathbf{G}$ is yet to be determined. Hence, evaluate (2.8) at $R$, solve the resulting expression for $\boldsymbol{\Psi}_\varepsilon^-$ then insert this boundary intensity into (2.8) and rearrange to obtain:

$$\boldsymbol{\Psi}_r^- = \rho^{-2}\widetilde{\mathbf{Z}}_{\varepsilon:r}\mathbf{Z}_{\varepsilon:R}^{-1}\boldsymbol{\Psi}_R^- + \rho^{-2}\widetilde{\mathbf{Z}}_{\varepsilon:r}\mathbf{Z}_{\varepsilon:R}^{-1}\mathbf{Q}_{R:\varepsilon}^-\boldsymbol{\lambda} - \mathbf{Q}_{r:\varepsilon}^-\boldsymbol{\lambda} \qquad (2.11)$$

where $\rho = r/R$ and $\widetilde{\mathbf{Z}}_{\varepsilon:r} = \mathbf{Z}_{\varepsilon:r}\mathbf{G}$. We resolve $\widetilde{\mathbf{Z}}_{\varepsilon:r}$ to satisfy the following conditions: (*a*) boundary condition at $r = R$, (*b*) reciprocity of path, and (*c*) symmetry at the center of the sphere. When these conditions are satisfied, the solution is necessarily conservative. There is no need to impose the conservation of energy a condition on the solution, it is already embedded into the structure of the discrete set of ordinates. The boundary condition is satisfied if:

$$\widetilde{\mathbf{Z}}_{\varepsilon:R}\mathbf{Z}_{\varepsilon:R}^{-1} = \mathbf{I} \quad \Rightarrow \quad \widetilde{\mathbf{Z}}_{\varepsilon:r} = \widetilde{\mathbf{U}}_{\varepsilon:r}\left[\alpha^{\beta_n^n} e^{\lambda_n(r-\varepsilon)}\right]\widetilde{\mathbf{U}}_{\varepsilon:r}^{-1} \qquad (2.12)$$

We find $\alpha$ by forcing (2.11) to satisfy the reciprocity of path, i.e., vacuum condition, and simultaneously the boundary condition at the surface of the sphere. Keeping in mind that $\alpha^{\beta_n^n} = e^{\beta_n^n \ln\alpha}$, then $\widetilde{\mathbf{U}}_{\varepsilon:r}$ differs from $\mathbf{U}_{\varepsilon:r}$ only in $\ln\alpha$ expressed as $\ln\zeta$ in $\Gamma$ of (2.7). It follows that $\widetilde{\mathbf{U}}_{\varepsilon:r}^{vac} = \widetilde{\mathbf{U}}_{\varepsilon:\varepsilon} = \mathbf{V}$ and (2.11) yields:

$$\boldsymbol{\Psi}_r^- = \rho^{-2}\mathbf{V}\left[\alpha^{\beta_n^n}\right]\left[\zeta_R^{-\beta_n^n}\right]\mathbf{V}^{-1}\boldsymbol{\Psi}_R^- \qquad (2.13)$$

Finiteness of the intensity $\forall r$ and $\forall N$, and application of the surface boundary condition to this expression yield the value $\alpha = \zeta_R$. Hence:

$$\widetilde{\mathbf{Z}}_{\varepsilon:r}\mathbf{Z}_{\varepsilon:R}^{-1} = \widetilde{\mathbf{U}}_{\varepsilon:r}\left[\zeta_R^{\beta_n^n} e^{\lambda_n(r-\varepsilon)}\right]\widetilde{\mathbf{U}}_{\varepsilon:r}^{-1}\mathbf{U}_{\varepsilon:R}\left[\zeta_R^{-\beta_n^n} e^{-\lambda_n(R-\varepsilon)}\right]\mathbf{U}_{\varepsilon:R}^{-1} \qquad (2.14)$$

and with that the solution of the transfer equation for the centripetal intensity is complete.

## 2.3. The centrifugal intensity with natural boundary condition

Unlike the centripetal solution, the centrifugal solution (2.4) can be completed with either a reflective boundary condition at the surface of the pellet, or an *albedo* boundary condition at the interface with the pellet. Both of the boundary conditions are ultimately expressed in terms of the natural boundary condition $\boldsymbol{\Psi}_R^-$.

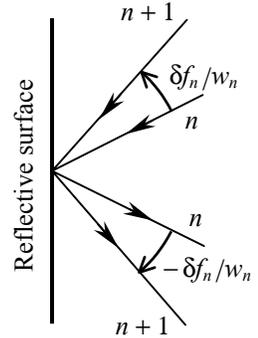

**Figure 1.** Symmetry effect on the angular derivative upon specular reflection.

(*a*) *Reflective boundary condition*. We cannot equate $\boldsymbol{\Psi}_\varepsilon^+$ with $\boldsymbol{\Psi}_\varepsilon^-$, instead we seek and expression for $\boldsymbol{\Psi}_\varepsilon^+$ in terms of $\boldsymbol{\Psi}_\varepsilon^-$. The reason we cannot equate these two intensities is justified by the fact that the angular derivative changes sign upon specular reflection (fig. 1). Such a change in sign is tantamount to application of symmetry on $\boldsymbol{\Psi}_\varepsilon^-$. Let $\mathbf{H}$ be the needed symmetry operator, yet to be determined, such that $\boldsymbol{\Psi}_\varepsilon^+ = \mathbf{H}\boldsymbol{\Psi}_\varepsilon^-$. We first obtain the centripetal intensity at the surface of the pellet by evaluating (2.11) at $r = \varepsilon$:

$$\boldsymbol{\Psi}_\varepsilon^- = \zeta_R^{-2}\widetilde{\mathbf{Z}}_{\varepsilon:\varepsilon}\mathbf{Z}_{\varepsilon:R}^{-1}\boldsymbol{\Psi}_R^- + \zeta_R^{-2}\widetilde{\mathbf{Z}}_{\varepsilon:\varepsilon}\mathbf{Z}_{\varepsilon:R}^{-1}\mathbf{Q}_{R:\varepsilon}^-\boldsymbol{\lambda} \qquad (2.15)$$

Operate $\mathbf{H}$ on (2.15), insert the resulting expression into (2.4), define $\widetilde{\mathbf{S}}_{\varepsilon:r} = \mathbf{S}_{\varepsilon:r}\mathbf{H}\widetilde{\mathbf{Z}}_{\varepsilon:\varepsilon}$ and con-strain the resulting expression to the reciprocity of path to obtain the complete solution for the centrifugal intensity in terms of the natural boundary condition:

$$\Psi_r^+ = \rho^{-2}\,\widetilde{\mathbf{S}}_{\varepsilon:r}\,\mathbf{Z}_{\varepsilon:R}^{-1}\,\Psi_R^- + \rho^{-2}\,\widetilde{\mathbf{S}}_{\varepsilon:r}\,\mathbf{Z}_{\varepsilon:R}^{-1}\,\mathbf{Q}_{R:\varepsilon}^-\,\lambda + \mathbf{Q}_{r:\varepsilon}^+\,\lambda \qquad (2.16)$$

$$\widetilde{\mathbf{S}}_{\varepsilon:r} = \widetilde{\mathbf{V}}_{\varepsilon:r}\left[\zeta_R^{\beta_n^n}\,e^{-\lambda_n(r-\varepsilon)}\right]\widetilde{\mathbf{V}}_{\varepsilon:r}^{-1}$$

where $\widetilde{\mathbf{V}}_{\varepsilon:r}$ is obtained from $\mathbf{V}_{\varepsilon:r}$ by replacing $\ln\zeta$ of (2.7) by $\ln\alpha$. One can verify that in vacuum, (2.11) and (2.16) become identically the same satisfying rigorously the inverse square law:

$$\Psi_r^+ = \Psi_r^- = \frac{1}{r^2}\left(R^2\,\Psi_R^-\right) \qquad (2.17)$$

This result is expected since (2.2) and (2.3) become identically the same Euler first order equation in matrix form [1].

Another mathematical property of (2.11) and (2.16) is that they allow only the normal intensity to propagate deep into the sphere all the way through the center. All the intensities from all other directions vanish at the center; the normal intensity in $S_{2N}$ is nominally represented by the $N$th direction for which $\beta_N^N = 0$.

(*b*) *Albedo boundary condition*. Evaluate (2.16) at $R$ and rearrange to obtain:

$$\Psi_R^+ = \mathbf{A}_{\varepsilon:R}\,\Psi_R^- + \mathbf{A}_{\varepsilon:R}\,\mathbf{Q}_{R:\varepsilon}^-\,\lambda + \mathbf{Q}_{R:\varepsilon}^+\,\lambda \qquad (2.18)$$

$$\mathbf{A}_{\varepsilon:R} = \widetilde{\mathbf{S}}_{\varepsilon:R}\,\mathbf{Z}_{\varepsilon:R}^{-1}$$

It is apparent that the nature of the operation of $\mathbf{A}_{\varepsilon:R}$ on its operand adheres the significance of *albedo* to this operator. This albedo has little in common with the traditional albedo defined as the ratio of the incident to returned radiation resulting from scattering in the medium. The albedo in the present case is simply a manifestation of symmetry, a geometric property of the volume, whether or not there is scattering.

We observe that the radius $R$ of (2.18) can be treated as an independent parameter and eventually an independent variable. Choose an infinitesimal radius $\xi$ and set $R = \varepsilon$, and let the physical parameters of (2.18) be those of the pellet. Then the centrifugal intensity at the surface of the pellet becomes:

$$\Psi_\varepsilon^+ = \mathbf{A}_{\xi:\varepsilon}\,\Psi_\varepsilon^- + \mathbf{A}_{\xi:\varepsilon}\,\mathbf{Q}_{\varepsilon:\xi}^-\,\lambda + \mathbf{Q}_{\varepsilon:\xi}^+\,\lambda \qquad (2.19)$$

Clearly, if $\Psi_\varepsilon^-$ is evaluated from (2.11), then (2.19) can serve as a boundary condition to (2.4), an *albedo* boundary condition to distinguish it from the specular reflection boundary condition.

*2.4. The heterogeneous sphere*

The solutions developed in the previous sections are valid for a homogeneous medium. If the sphere is not homogeneous, then the diagonal elements of the $\Gamma$-matrix become:

$$\Gamma_n^n = \left(2+\beta_n^n\right)\ln\zeta - L_n(\varepsilon,r) \qquad (2.20)$$

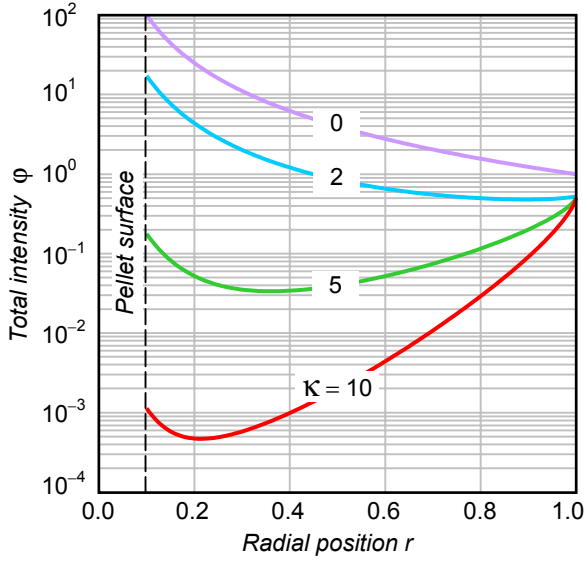
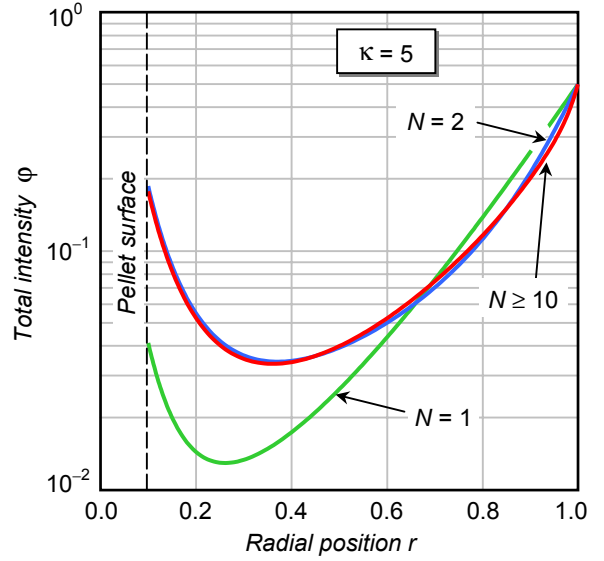

**Figure 2.** Graphs of total intensities in spheres of outer radius $R = 1$ and of different opacities embedding a pellet of radius $\varepsilon = 0.1$. The spheres are exposed to external isotropic radiation field of normalized intensity and specular reflection is applied at the surface of the pellet.

**Figure 3.** Graphs of the total intensity in a sphere of outer radius $R = 1$ embedding a pellet of radius $\varepsilon = 0.1$ for various orders $N$ of discrete ordinates. The sphere is exposed to external isotropic radiation field of normalized intensity with reflective boundary condition at the surface of the pellet.

$$L_n(\varepsilon, r) = \int_\varepsilon^r \lambda_n \mathrm{d}r \qquad (2.21)$$

where $L_n(\varepsilon, r)$ is the optical thickness of the spherical shell, to be used in $\mathbf{S}_{\varepsilon \cdot r}$; the minus sign of (2.20) is used for the solution of the centrifugal intensity, it should be reversed to a plus sign for the solution of the centripetal intensity. It follows that the solutions of the previous sections apply to non-homogeneous spheres by simply making the substitutions: $\lambda_n(r - \varepsilon) \leftarrow L_n(\varepsilon, r)$.

If the sphere is stratified into $M$ homogeneous strata, and each $m$th stratum is characterized by a uniform opacity $\kappa_m$ and geometric thickness $\Delta r_m$, then the optical thickness of a sphere defined by its outer radius $r_k : k \leq M$, is obtained by the summation:

$$L_{n,k} = \sum_{m=1}^{m=k} \lambda_{n,m} \Delta r_m \qquad (2.22)$$

## 3. Quantitative analysis

We considered graphical and numerical data to demonstrate the native characteristics of the solutions using the proposed set of discrete ordinates. We experimented with pure absorbers homogeneous spheres of radius $R = 1$ and a pellet of radius $\varepsilon = 0.1$ with reflective surface. The expressions (2.11) and (2.16) are used in all calculations using 16 significant digits arithmetics (16-SDA). The spheres are immersed in uniform and isotropic radiation field of normalized intensity $\overline{\psi}_R = 1$.

Graphs of total intensities are shown in figure 2. The graph of the intensity in vacuum, desig-nated by the opacity $\kappa = 0$, obeys rigorously the inverse square law (2.17), although it is generated with (2.11) and (2.16). This result validates numerically the reciprocity of path for all orders of discrete ordinates. The other graphs of figure 2 are generated with $N = 10$. They show a minimum of intensity at some point in the interior of the sphere as predicted from the centripetal equation (2.3). The complexity of the analytical expressions, however, makes it impossible to determine the value of the radial point where the minimum occurs.

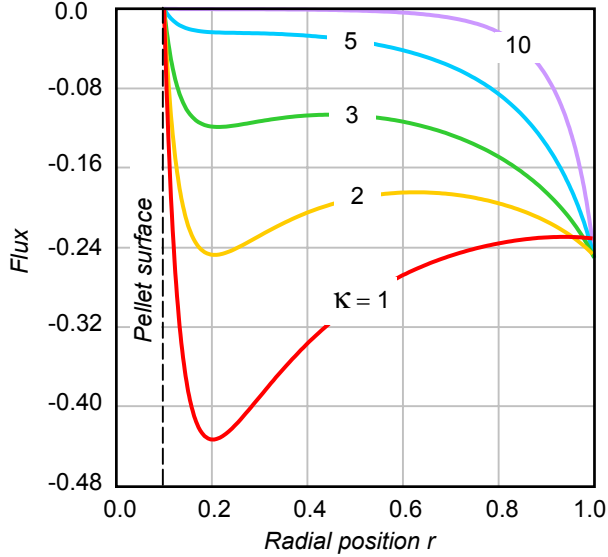
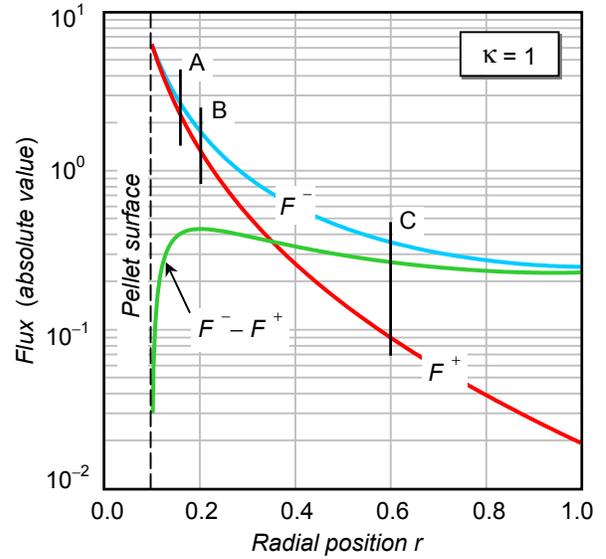

**Figure 4.** Graphs of fluxes in spheres of outer radius $R = 1$ and of different opacities, embedding a pellet of radius $\varepsilon = 0.1$. The spheres are exposed to external isotropic radiation field of normalized intensity and specular reflection is applied at the surface of the pellet.

**Figure 5.** Graphs of centripetal flux $F^-$ and centrifugal flux $F^+$, and their difference $F$, in a sphere of radius $R = 1$ embedding a pellet of radius $\varepsilon = 0.1$; reflective boundary condition is imposed on the surface of the pellet. The difference between the partial fluxes is larger at B than it is at A or at C, which explains the spike of the flux in the proximity of the surface of the pellet.

It is worth noting that the graphs of figure 2 are quite different from the graphs one would obtain using the solution of the integral equation in a solid homogeneous sphere. Specifically [3]:

$$\varphi_r = \frac{1}{2}\int_{-1}^{+1} e^{-\kappa s(r,\mu)}\psi_R^- \, d\mu \qquad (2.23)$$

$$s = r\mu + \sqrt{R^2 - r^2\eta^2}$$

Using our isotropic boundary intensity, (2.23) yields a flat total intensity everywhere in the cavity, quite a contrast with the $1/r^2$ of the present solution. This comparison shows that the integral equations (IE) and the ingtro-differential equation (IDE) are not equivalent for the total intensity in the cavity. This sounds paradoxical since the solution of IE, which is the integrand of (2.23), and the solution of (IDE) given as (2.11) and (2.16), both of them satisfy the same boundary condition at the surface of the sphere, the reciprocity of path and the symmetry at the center.

The transfer equation defines a boundary value problem and it has a unique solution for each set of boundary conditions. Therefore, IE and IDE should be construed as two different boundary value problems in the interior of spheres. We realize that IDE yields a solution for the normal incidence intensity at the surface compatible with the empirical $1/r^2$ law, the solution of IE is flat everywhere in the sphere. We conclude from this observation that IE defines a boundary value problem not representative of energy flow in the interior of spheres. Hence, IE is not equivalent to IDE in spheres. It was shown that IE and IDE are equivalent in spherical geometry in the case of centrifugal one-stream radiation field [1].

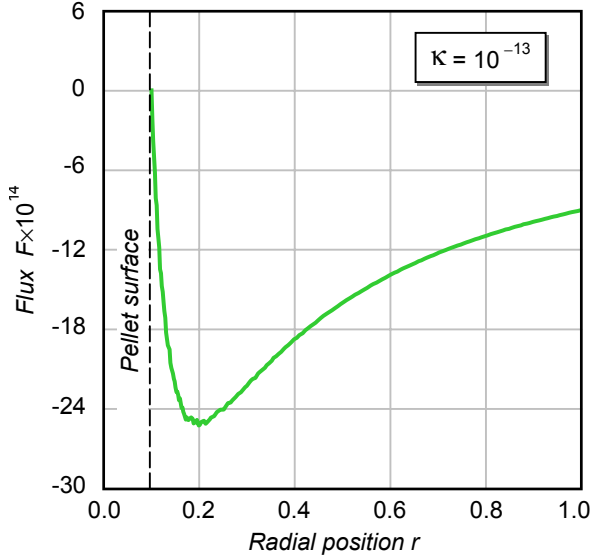
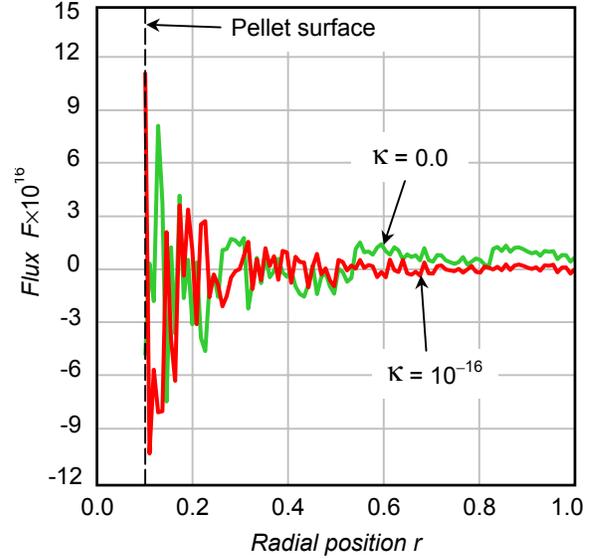

**Figure 6.** Graph of the flux in a sphere of outer radius $R = 1$ and opacity $\kappa = 10^{-13}$ embedding a pellet of radius $\varepsilon = 0.1$. The sphere is exposed to external isotropic radiation field of normalized intensity and specular reflection is applied at the surface of the pellet.

**Figure 7.** Numerical fluctuations of the flux about zero in vacuum space and in a medium that comports as a vacuum in finite arithmetics of 16 significant digits.

The graphs of figure 3 illustrate visually the sensitivity of the total intensity on the order $N$ of discrete ordinates. Increasing $N$ above 10 makes no visual difference on the graph, understandably in pure absorber.

Graphs of fluxes in absorbing spheres are also generated and are shown in figure 4. It is apparent that some fluxes reach a minimum before they spike and then vanish at the surface of the pellet; the flux vanishes at the surface of the pellet by reason of specular reflection boundary condition. Some fluxes in figure 4 do not spike. Our experimentation revealed that the fluxes that do not spike in the case of the sphere of figure 5, do spike if the pellet radius is smaller than 0.1.

The spike is not a numerical aberration it is real. Figure 5 shows graphically where the spike comes from. For a given sphere, there is a range of opacities for which the difference between the centrifugal component $F^+$ and the magnitude of the centripetal component $F^-$ of the flux has a maximum at some point, B in figure 4. The occurrence of that maximum and its radial position depend on $R$, $\varepsilon$ and $\kappa$. It was not possible to derive an analytical expression for that radial position, nor was it possible to determine for what opacity it occurs. The location of the spikes of figure 4 suggests as if they occur at $2\varepsilon$. This happens to be the case in this sphere and in many

**Table 1.** Numerical values for the total intensities $\varphi_0$ at the surface of the pellet, $\varphi_R$ at the surface of the sphere, the flux $F_R$ at the surface of the sphere, and the conservation of energy $E$. These data are generated with different orders $N$ of discrete ordinates in spheres having the same radii $R = 1$ and $\varepsilon = 0.1$ but different opacities.

| | $\kappa = 0.1$ | | | | $\kappa = 5.0$ | | | |
|---|---|---|---|---|---|---|---|---|
| $N$ | $\varphi_0$ | $\varphi_R$ | $F_R$ | $E \times 10^{16}$ | $\varphi_0$ | $\varphi_R$ | $F_R$ | $E \times 10^{11}$ |
| 1 | 85.5658 | 0.866075 | −0.0773215 | −1.11 | 0.041211 | 0.500000 | −0.288675 | 0.57 |
| 2 | 81.3804 | 0.833155 | −0.0738217 | 5.00 | 0.187153 | 0.500005 | −0.260629 | 1.16 |
| 4 | 77.7177 | 0.810003 | −0.0709241 | 4.58 | 0.177692 | 0.500006 | −0.252877 | 2.35 |
| 8 | 75.2199 | 0.800208 | −0.0691065 | 6.25 | 0.177869 | 0.500006 | −0.250753 | 4.73 |
| 16 | 74.1778 | 0.799348 | −0.0683759 | 12.1 | 0.177869 | 0.500006 | −0.250190 | 9.48 |
| 32 | 74.0831 | 0.799759 | −0.0682084 | 19.2 | 0.177869 | 0.500006 | −0.250044 | 18.9 |

other ones, but it is not the case in yet many other spheres.

Numerical irregularities of the flux manifest in spheres with very small opacities, $\kappa < 10^{-13}$, including in vacuum. These irregularities are attributable to numerical round off. Figure 6 shows the threshold opacity in 16-SDA for which the effect of the round off becomes mostly noticeable at the spike location. The effect of the round off with $\kappa = 10^{-16}$ and in vacuum, $\kappa = 0.0$, are shown in figure 7; the data point at $r = R$ in vacuum is exactly $F_R = 0.0$.

Numerical data were also analyzed in order to determine the convergence of the total intensity and the flux with increasing order of discrete ordinates. A sample of these data is shown in Table 1 for two spheres of different opacities. The total intensity at the surface of the sphere, designated by $\varphi_R$, appears to converge faster than the intensity at the surface of the pellet designated by $\varphi_0$. This is expected in a pure absorbing medium. The principal component of the intensity at the surface of the sphere comes from the boundary condition. Not much of centrifugal intensity contributes to the total intensity at the surface of the sphere, particularly in strongly absorbing media.

The flux at the surface of the pellet is not reported in Table 1. It is $F_0 = 0.0$ by reason of specular reflection. The flux at the surface of the sphere is reported as $F_R$. It converges smoothly. The conservation of energy designated by $E = R^2 F_R - \varepsilon^2 F_0 + \kappa \varphi_{int}$ in the table is obtained by integrating the transfer equation over all directions and over the volume of the spherical shell bounded by $\varepsilon$ and $R$; the quantity $\varphi_{int}$ is the volume integral of the total intensity. Ideally, $E$ must be zero. It is apparent from Table 1 that $E$ is not quite zero, but very small. The discrepancy is attributable to round off: as $N$ increases the matrices become larger generating larger round off errors. Data in spherical cavity are not shown in the table. They are rigorously the values obtained in $1/r^2$ radiation field: $\varphi_0 = 100$, $\varphi_R = 1$ and $E = 0$ for all $N$ values we experimented with.

The highest order of the discrete ordinates in Table 1 is conspicuously no greater than $N = 32$. Larger values of $N$ returned overflow from the evaluation of the diagonal matrices. This is true in the case of $\kappa = 0.1$. In the case of $\kappa = 5$, $N = 40$ could be tolerated with noticeable round off particularly in $E$; stronger opacities tolerate greater values $N$.

## 4. Summary a conclusion

We obtained the analytic solution in *r* for the two-streams transfer equations in $S_{2N}$ representation. The formal solutions are straightforward with boundary conditions at the surface of the pellet. The novelty in the solution is in resolving the formal mathematical boundary conditions in terms of the natural boundary condition at the surface of the sphere. This was accomplished by taking the reciprocal of the spherical operator of the centripetal solution.

The proposed solution has a limited scope as expected to be the case with all analytic solutions. Nevertheless, it is a demonstration of the validity of the set of discrete ordinates weaved with un-normalized circular functions. This set enabled mapping the angular derivative of the differential calculus onto the discrete domain $S_{2N}$.

The solutions obtained with this approach satisfy the necessary conditions descriptive of the flow of energy in spheres, specifically, the natural boundary condition at the surface of the sphere, the symmetry condition at the center of the sphere, the reciprocity of path, the inverse square law and the conservation of energy. Traditionally, the solution in discrete ordinates formalism requires a starter specific intensity. A starter intensity is not needed in this approach nor is it required to impose a condition on the nature of the intensity or on its spatial derivative at the center of the sphere.

The upper limit of the order of discrete ordinates that can be used without overflow messages is determined by the limitation of the finite arithmetics of the computing machine in the evaluation of the exponential of $\beta_n^n \ln \zeta_R + \lambda(r-\varepsilon)$. Nevertheless, the tabulated data demonstrate the potential of the proposed discrete ordinates method to improve on the limitation of the finite arithmetics by developing an adequate numerical iterative algorithm.

We have not displayed data for the intensities in spheres with emission. The expressions for the emission terms in the analytic solutions are quite unfriendly on digital computers, save the unbearable low speed of computations. The round off errors are too great to merit any value to the output. This problem, however, is easier to deal by using adequate numerical methods.

**Acknowledgment**. This work was funded by the Department of Navy, grant N00173-001-G010.